\documentclass[%
 aip,
 amsmath,amssymb,
 citeautoscript, 
 reprint,%
]{revtex4-1}

\usepackage{xcolor}

\newcommand{\blue}[1]{{#1}}

\usepackage{graphicx}
\usepackage{dcolumn}
\usepackage{bm}

\usepackage[utf8]{inputenc}
\usepackage[T1]{fontenc}
\usepackage{mathptmx}
\usepackage{etoolbox}

\usepackage{fancyhdr}

\makeatletter
\def\@email#1#2{%
 \endgroup
 \patchcmd{\titleblock@produce}
  {\frontmatter@RRAPformat}
  {\frontmatter@RRAPformat{\produce@RRAP{*#1\href{mailto:#2}{#2}}}\frontmatter@RRAPformat}
  {}{}
}%
\makeatother

\begin{document}


\title[]{Experimental demonstration of ultrathin broken-symmetry metasurfaces with controllably sharp resonant response}

\author{Odysseas Tsilipakos}
\email{otsilipakos@iesl.forth.gr}
\affiliation{Institute of Electronic Structure and Laser, Foundation for Research and Technology-Hellas (FORTH-IESL), GR-70013 Heraklion, Crete, Greece}

\author{Luca Maiolo}
\author{Francesco Maita}
\author{Romeo Beccherelli}
\affiliation{Consiglio Nazionale delle Ricerche, Istituto per la Microelettronica e Microsistemi (CNR-IMM), Roma 00133, Italy}

\author{Maria Kafesaki}
\affiliation{Institute of Electronic Structure and Laser, Foundation for Research and Technology-Hellas (FORTH-IESL), GR-70013 Heraklion, Crete, Greece}
\affiliation{Department of Materials Science and Technology, University of Crete, GR-70013 Heraklion, Crete, Greece}

\author{Emmanouil E. Kriezis}
\author{Traianos V. Yioultsis}
\affiliation{School of Electrical and Computer Engineering, Aristotle University of Thessaloniki (AUTH), GR-54124 Thessaloniki, Greece}

\author{Dimitrios C. Zografopoulos}
\affiliation{Consiglio Nazionale delle Ricerche, Istituto per la Microelettronica e Microsistemi (CNR-IMM), Roma 00133, Italy}

\date{\today}



\begin{abstract}
    Symmetry-protected resonances can be made to couple with free space by introducing a small degree of geometric asymmetry, leading to controllably-sharp spectral response. Here, we experimentally demonstrate a broken-symmetry metasurface for the technologically important low millimeter wave spectrum. The proposed metasurface is fabricated on an ultrathin polyimide substrate, resulting in a low loss and flexible structure. Measurements inside an anechoic chamber experimentally verify the theoretically predicted sharp spectral features corresponding to quality factors of several hundreds. The demonstrated sharp response is also observed with the complementary structure which responds to the orthogonal linear polarization (Babinet's principle). The designed metasurfaces can be exploited in diverse applications favoured by a controllably-sharp spectral response, e.g., filtering, sensing, switching, nonlinear applications, in either reflection or transmission mode operation. More generally, the demonstrated fabrication process provides a generic platform for low-cost, large-scale engineering of metasurfaces with minimal substrate-induced effects.
\end{abstract}

\maketitle

\thispagestyle{fancy}
\lhead{The following article has been accepted by Applied Physics Letters. After it is published, it will be found at https://doi.org/10.1063/5.0073803}




Metasurfaces (MSs), planar periodic structures composed of a single layer of meta-atoms, have attracted considerable interest in recent years. Despite being quite thin compared to volumetric metamaterials, metasurfaces can interact strongly with impinging electromagnetic radiation by virtue of their resonant nature. As a result, they have been investigated for an abundance of applications \cite{Chen:2016,Sun:2019,Tsilipakos2020}, among which perfect absorption, polarization control, (achromatic) wavefront manipulation, frequency generation, and lasing.

Metasurfaces exhibiting sharp resonances (high quality factors) are associated with strong field enhancement, long energy storage, and enhanced light matter interaction; as a result, they are particularly interesting for filtering, sensing, switching, nonlinear, and lasing applications. Recently, metasurfaces supporting resonances with diverging (theoretically to infinity) quality factors are being intensively investigated in the context of dark, trapped, symmetry-protected, anapole and BIC (bound states in the continuum) resonances \cite{Campione2016,He2018,Hsu2016,Koshelev2018,Karl2019,Algorri_2021}. The common underlying principle is that an initially non-radiative resonance is made to controllably couple with free space; at the onset of this coupling the associated \emph{radiation} quality factor ($Q_\mathrm{rad}$) diverges to infinity. An important consideration in the case of almost non-radiative (quasi-dark) resonances is that the maximum attainable \emph{total} quality factor ($Q_\mathrm{tot}$) will be bound by resistive losses. In fact, the darker the resonance is (higher $Q_\mathrm{rad}$), the more confined the fields are to the meta-atom leading to higher absorption in the materials (Ohmic losses). Thus, in the interest of obtaining high total quality factors, it is essential not only to start with a high radiation quality factor, but also select the physical structure (material system) to be adequately low loss.

\begin{figure*}
\centering
\includegraphics[width=17.8cm]{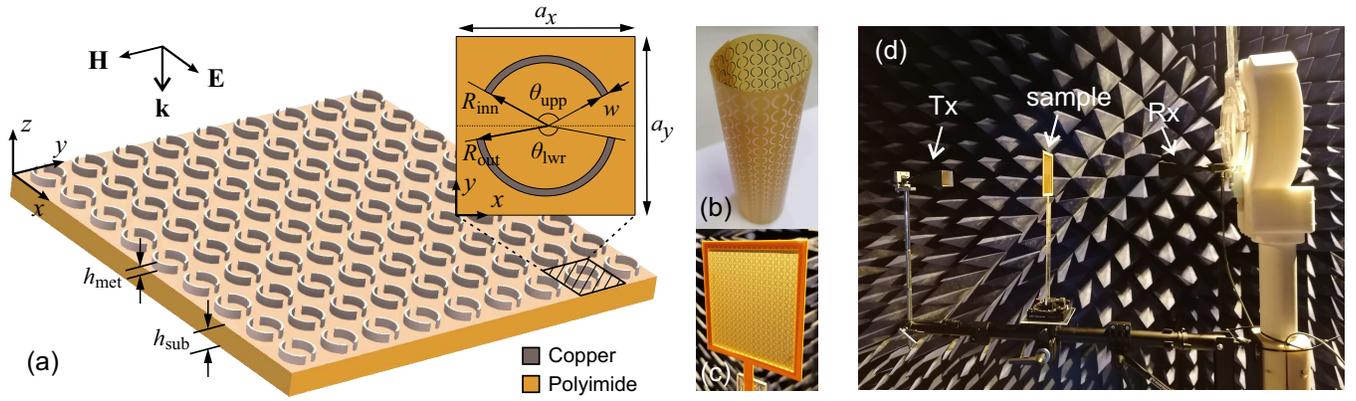}
\caption{(a)~Schematic of proposed ultrathin metasurface with annular metal sectors; metallization (9~$\mu$m) and substrate (25~$\mu$m) thicknesses are shown in scale. The unit cell with annotated geometrical parameters is included as an inset. (b)~Fabricated metasurface ($19\times 19$ unit cells) demonstrating significant flexibility due to the ultrathin substrate. (c)~Sample mounted on a 3D-printed frame. (d)~Measurement setup with transmitting (Tx) and receiving (Rx) antenna inside an anechoic chamber.}
\label{fig:Structure}
\end{figure*}

Here, we employ a broken symmetry meta-atom as a straightforward strategy to implementing controllably bright (radiative) resonances. The proposed metasurface is depicted in Fig.~\ref{fig:Structure}. It consists of annular metallic (copper) sectors on top of a thin polyimide (PI) substrate. The two sectors are in general asymmetric $(\theta_\mathrm{upp}\neq\theta_\mathrm{lwr})$; the degree of asymmetry determines the radiation quality factor of the supported symmetry-protected resonance and, consequently, the linewidth of the associated spectral feature. Such a geometry was first proposed by Fedotov et al. \cite{Fedotov2007} on standard printed circuit board (PCB) technology for operation at $\sim6$~GHz. Here, we redesign the structure for operation at the the bottom of the millimeter wave band ($\sim25$~GHz), which has received a lot of interest lately in connection with 5G communications. Importantly, we adopt an ultrathin ($h_\mathrm{sub}=25~\mu$m) low-cost polyimide substrate, much thinner than typical PCB substrates, resulting in reduced Ohmic losses and the ability to support higher resistive quality factors. The use of such deeply subwavelength substrates can almost reproduce the performance of free-standing structures, minimizing substrate-induced detrimental effects \cite{Maiolo2019}. At the same time, it provides mechanical stability and can be used for metasurfaces with isolated meta-atoms not possible in free-standing configurations. The resulting metasurface is quite flexible [Fig.~\ref{fig:Structure}(b)] and can conformally coat objects or textiles. (Additional photos of the fabricated samples can be found in  Fig.~S1 of the Supplementary Material.) Substrate ($h_\mathrm{sub}=25~\mu$m) and metallization ($h_\mathrm{met}=9~\mu$m) layers are shown in scale in the schematic of Fig.~\ref{fig:Structure}(a). Thanks to the employed ultrathin-substrate approach, we are able to numerically calculate (detailed plane-wave scattering and eigenvalue simulations) and, more importantly, experimentally demonstrate $Q_\mathrm{tot}$ values in the order of a few hundred, namely one order of magnitude higher than earlier measurements on this type of metasurfaces \cite{Fedotov2007}. The study is also extended to the complementary structure (annular slots on a metallic film) which possesses an entirely similar response for the orthogonal linear polarization with reflection and transmission coefficients reversed, as dictated by Babinet's principle. Due to the ultra-thin metallization and substrate layers, the Babinet principle is found to hold very well apart from a shift in the resonant frequencies, as will be demonstrated.

The samples investigated in this work were manufactured in the CNR-IMM labs based on a photolithographic process flow. First, lithographic masters (masks) were designed and fabricated in order to pattern squared samples of size $12~\mathrm{cm}~\times~12~\mathrm{cm}$. Then, a commercial low-cost polyimide substrate with copper foil (Upisel-N copper-clad laminate by UBE Industries), a standard material for flexible circuits, was adopted and the dry resin FP415 by ElgaEurope was deposited on the substrates using a low temperature ($120^{\circ}$C) lamination technique (Bungard RLM 419P). Subsequently, the samples were patterned by using an EVG610 mask aligner and copper was etched through a wet process in a spray-etching machine (RotaSpray Plus), controlling possible overetching. 
Then, the remaining resin was removed and the samples were rinsed with deionized water, drying gently the metasurfaces in nitrogen flux. Finally, the flexible samples were properly cut and adapted to a 3D-printed acrylonitrile butadiene styrene (ABS) frame of $1$~mm thickness to allow handling during the measurements [Fig.~\ref{fig:Structure}(c)]. In all cases, copper overetching was below $10~\mu$m, as verified by visual inspection of the samples under an optical microscope.

The measurement setup is shown in Fig.~\ref{fig:Structure}(d). The metasurface sample, mounted on the ABS frame, is positioned between a pair of transmitting (Tx) and receiving (Rx) K-band (17.6–26.7 GHz) horn antennas (FLANN Microwave 20240) with a standard gain  of 20~dBi, which are connected to a vector network analyzer (Anritsu 37397D). All measurements have been conducted inside an anechoic chamber. \blue{Before conducting the measurements, a transmission ($S_{21}$) response calibration was performed with the sample and holder (frame) missing, i.e., the case of a free-space link between transmitting and receiving antennas. Note that the distances between antennas and sample are sufficiently larger than the near-field limit of the antennas.}

We first perform plane-wave scattering numerical simulations ($x-$polarized incidence) for the metasurface in Fig.~\ref{fig:Structure}(a) using the commercial software COMSOL Multiphysics which implements the vectorial Finite Element Method. A single unit cell is simulated in the frequency domain with periodic boundary conditions on the $x-$ and $y-$boundaries. The geometrical properties with reference to the inset of Fig.~\ref{fig:Structure}(a) are $a_x=a_y=5.2$~mm, $R_\mathrm{out}=2.25$~mm, $R_\mathrm{inn}=1.95$~mm, $w=R_\mathrm{out}-R_\mathrm{inn}=0.3$~mm, $h_\mathrm{sub}=25~\mu$m, $h_\mathrm{met}=9~\mu$m, $\theta_\mathrm{lwr}=160$~deg, and $\theta_\mathrm{upp}$ is kept as a free parameter determining the level of structural asymmetry $\gamma=(\theta_\mathrm{lwr}-\theta_\mathrm{upp})/\theta_\mathrm{lwr}$. The material parameters are $\varepsilon_r^\mathrm{PI}=3.2(1-j0.006)$ and $\sigma_\mathrm{Cu}=5.8\times 10^7$~S/m.

\begin{figure*}
\centering
\includegraphics[width=17.8cm]{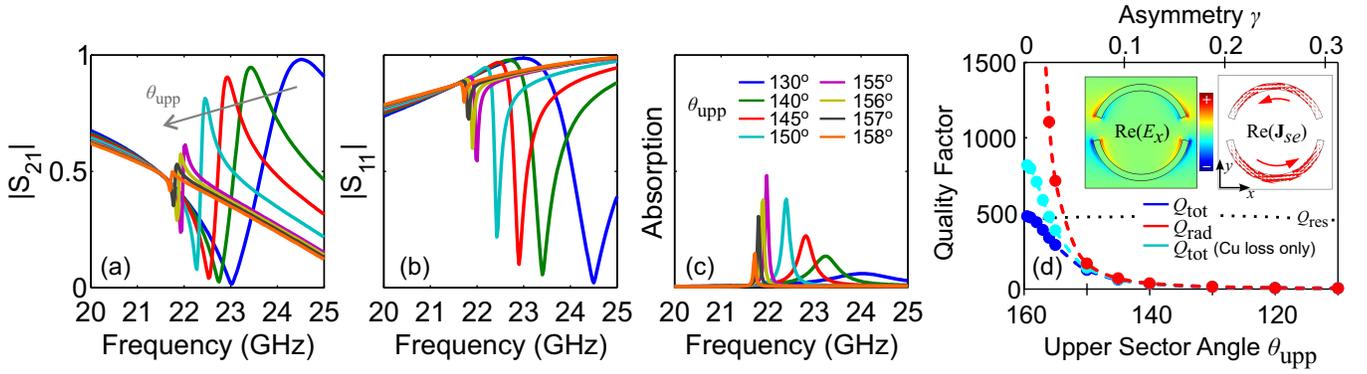}
\caption{Simulations of plane-wave scattering ($x$-polarized incidence) from a metasurface with different $\theta_\mathrm{upp}$ values resulting in different degrees of asymmetry ($\theta_\mathrm{lwr}=160$~deg throughout). (a)~Co-polarized transmission amplitude coefficient ($|S_{21}|$), (b)~co-polarized reflection amplitude coefficient ($|S_{11}|$) and (c)~absorption. (d)~Total ($Q_\mathrm{tot}$) and radiation ($Q_\mathrm{rad}$) quality factors as a function of $\theta_\mathrm{upp}$. $Q_\mathrm{rad}$ diverges to infinity for a symmetric meta-atom. $Q_\mathrm{tot}$ remains finite due to the presence of resistive losses ($Q_\mathrm{res}\sim 500$). If absorption in the substrate is zeroed out (e.g. with a smaller thickness and/or lower loss), $Q_\mathrm{tot}$ can reach $\sim 850$. Insets depict the mode profile ($E_x$ field component and surface current $\mathbf{J}_{se}$) for the case  $\theta_\mathrm{upp}=145$~deg.}
\label{fig:Simulations}
\end{figure*}

The magnitude of $S_{21}$ (co-polarized transmission amplitude coefficient) and $S_{11}$ (co-polarized reflection amplitude coefficient) are depicted in Fig.~\ref{fig:Simulations}(a) and (b), respectively, for different values of $\theta_\mathrm{upp}$. The meta-atom is most asymmetric for $\theta_\mathrm{upp}=130$~deg. The spectral feature in this case is widest since the supported resonance is brightest (most radiative).
As $\theta_\mathrm{upp}$ increases and approaches $\theta_\mathrm{lwr}$, the degree of asymmetry decreases; the supported resonance becomes sharper as testified by the narrower linewidth of the spectral feature. At the same time it becomes weaker, since the resonance is not as effectively excited, until it finally disappears for a symmetric meta-atom. The Fano lineshape evident in Fig.~\ref{fig:Simulations}(a), (b) is a result of interference between the resonance under study (fast process), a low-$Q$ mode possessing a higher resonant frequency ($>25$~GHz) and the direct pathway via the continuum comprising non-resonant contributions at the frequencies under study. For details see the Supplementary Material (Sections S2 and S3). Figure~\ref{fig:Simulations}(c) depicts the absorption (power coefficient), as calculated by integrating the power loss density inside both metal and substrate regions. This approach is preferable to using just $A=1-|S_\mathrm{11}|^2-|S_\mathrm{21}|^2$, since such structures that do not possess $C_4$ rotational symmetry can exhibit strong cross-polarized response \cite{Shi2014}. Note that absorption is maximized for $\theta_\mathrm{upp}=155$~deg and reaches approximately 48\% (50\%  is the highest achievable with ultrathin single-metallization-layer metasurfaces that are solely electrically polarizable \cite{Fan2015}). For $\theta_\mathrm{upp}=155$~deg it approximately holds $Q\mathrm{rad}=Q\mathrm{res}\approx 500$ leading to critical coupling and maximum absorption \cite{Zhang2020,Ruan2010}. This can be verified in Fig.~\ref{fig:Simulations}(d) where we plot the total and radiation quality factors \cite{Christopoulos2019}, as calculated by eigenvalue simulations. The radiation quality factor (accounts for light leakage only) is calculated by momentarily zeroing out material losses and diverges as $\theta_\mathrm{upp}$ approaches $\theta_\mathrm{lwr}=160$~deg. For $\theta_\mathrm{upp}=158$~deg it can reach a value of $\sim 6\times 10^4$. However, the total quality factor is bound by $Q_\mathrm{res}\sim 500$. Material losses are dominated by losses in the copper metallization, but losses in PI contribute as well. By zeroing out loss in the substrate, the total quality factor can reach $850$ for $\theta_\mathrm{upp}=158$~deg. This limit can be approached by using even thinner custom PI substrates as thin as $8~\mu$m \cite{Maiolo2019}, at the expense of a non-standard manufacturing process and a reduction in mechanical stability.

The total quality factors depicted in Fig.~\ref{fig:Simulations}(d) are further corroborated by using an extended Fano formula to extract the quality factor by fitting to the simulated transmittance spectra $T=\left|S_{21}\right|^2$. In particular, the employed formula corresponds to a lossy Fano resonance \cite{Algorri_2021} (quasi-dark mode plus a direct non-resonant pathway) modulated by a Lorentzian resonance \cite{Chen_2014}, which capture the spectral features of the sharp symmetry-protected mode and the low-$Q$ bright mode, respectively (details in the Supplementary Material, Section S3). The results are summarized in Table~\ref{tab:Qs} and excellent agreement is observed.

\begin{table}
\centering
\caption{\label{tab:Qs} Comparison of total quality factor as calculated through an eigenvalue problem (depicted visually in Fig.~\ref{fig:Simulations}(d)) and extracted from the transmission spectra by fitting to an extended Fano formula, see Section S3 in the Supplementary Material.}
\begin{tabular}{lcccc}
\hline
$\theta_\mathrm{upp}$ (deg) & 130 & 140 & 145 & 150 \\
\hline
$Q_\mathrm{tot}$ (eigenvalue) & 14.3 & 34.9 & 61.8 & 126.4  \\
$Q_\mathrm{tot}$ (Fano fit) & 14.2 & 34.8 & 61.3 & 123.8 \\
\hline
$\theta_\mathrm{upp}$ (deg) & 155 & 156 & 157 & 158 \\
\hline
$Q_\mathrm{tot}$ (eigenvalue) & 292.0 & 339.5 & 391.6 & 441.5 \\
$Q_\mathrm{tot}$ (Fano fit) & 285.2 & 334.2 & 386.7 & 440.5 \\
\hline
\end{tabular}
\end{table}

The mode profile (eigenvector) of the resonance under study is depicted in the insets of Fig.~\ref{fig:Simulations}(d). The left inset depicts the $E_x$ field distribution (real part); \blue{it changes sign on either side of the $y=0$ plane, i.e., it is odd with respect to the $y$ axis.} When the meta-atom is symmetric ($\theta_\mathrm{upp}=\theta_\mathrm{lwr}$), the $E_x$ distribution is perfectly antisymmetric; as a result, the resonance cannot be excited by a normally-incident $x-$polarized plane wave. As soon as the meta-atom becomes asymmetric ($\theta_\mathrm{upp}\neq\theta_\mathrm{lwr}$), the distribution seizes to be perfectly antisymmetric and coupling with free space is allowed. The right inset depicts the induced electric surface current $\mathbf{J}_{se}$ (A/m) on the top metallization surface; an in-plane  resonant current loop is clearly observed. The two annular metal sectors can be viewed as curved cut-wires supporting fundamental electric dipole resonances (half a wavelength fitting in each metal sector). Each cut-wire is associated with an induced electric dipole moment $\mathbf{p}$. Since the two $p_x$ contributions are opposing each other, for a symmetric meta-atom the net result is zeroed out. When asymmetry is introduced, the counteracting contributions are not equal and the residual $p_x$ leads to coupling with the incident $E_x$ component.

\begin{figure}
\centering
\includegraphics{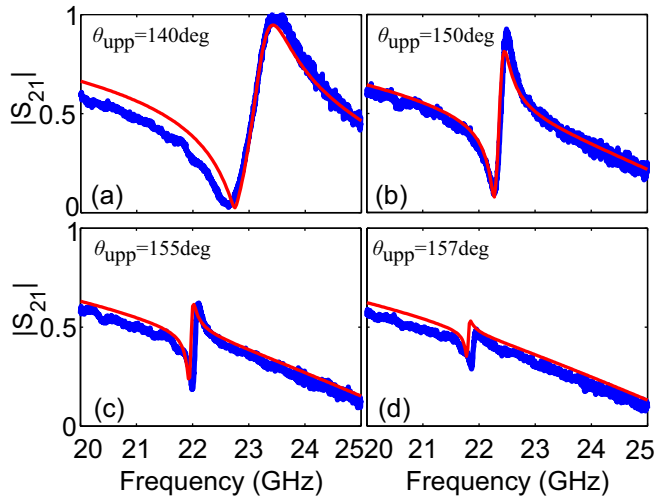}
\caption{Comparison of measured and simulated transmission coefficient ($|S_{21}|$) for the metasurface in Fig.~\ref{fig:Structure} with different upper-sector angles: (a)~$\theta_\mathrm{upp}=140$~deg,
(b)~$\theta_\mathrm{upp}=150$~deg, (c)~$\theta_\mathrm{upp}=155$~deg, (d)~$\theta_\mathrm{upp}=157$~deg
($\theta_\mathrm{lwr}=160$~deg throughout). For the remaining geometrical parameters see text. Excellent agreement is observed; the ultrathin low-loss substrate has allowed for obtaining sharp spectral features associated with high total quality factors.}
\label{fig:ExpVsSim}
\end{figure}

In Fig.~\ref{fig:ExpVsSim} measured transmission coefficients are plotted for different degrees of meta-atom asymmetry, $\theta_\mathrm{upp}=140,150,155,157$~deg, attempting to experimentally verify sharp spectral features and the associated high quality factors. In all four cases the agreement between measurement and simulation is excellent. Minor discrepancies are attributed to fabrication imperfections, as well as the effect of the finite metasurface sample and the non-planar incident wavefront. Importantly, the linewidth of the spectral features is reproduced very well, confirming that the ultrathin low-loss substrate has allowed for obtaining high total quality factors.

We now extend the study to the complementary (Babinet-inverted) structure, i.e., annular slots in an otherwise continuous copper film [see inset in Fig.~\ref{fig:Complementary}(a) and Fig. S5 in the Supplementary Material]. Due to the duality of Maxwell's equations and the Babinet principle \cite{Zhang2013}, we can expect that the complementary structure exhibits an entirely similar response with the initial one (annular metal sectors), when the incident polarization is the perpendicular one ($y-$ instead of $x-$polarized plane-wave incidence) and the forward/backward scattering directions are exchanged. The mode profile of the symmetry-protected resonance of the complementary structure can be found in the Supplementary Material (Fig.~S4).
In our structure the conditions for the Babinet principle hold reasonably well: the substrate and metallization layers are electrically ultrathin ($\lambda_0$ equals 13~mm at 23~GHz) and the cooper metallization is of very high conductivity. Indeed, this is verified in Fig.~\ref{fig:Complementary} where we plot transmission ($S_\mathrm{21}$) curves for the complementary structure ($y-$polarized incidence) and different values of $\theta_\mathrm{upp}$. Reflection ($S_\mathrm{11}$) curves for the initial structure ($x-$polarized incidence) first depicted in Fig.~\ref{fig:Simulations}(b) are overlaid here with dashed lines for convenience. They have been universally shifted by a frequency offset of +550~MHz. Apart from this $\sim2.5\%$ difference in the resonance frequencies, the Babinet principle is seen to hold well in our ultrathin structure. The response of the complementary structure has been also experimentally verified (see Fig.~S5 in the Supplementary Material), showing very good agreement with simulation results and an almost perfect reproduction of the spectral feature linewidth. Note that the complementary structure remains quite flexible despite the 9-$\mu$m-thick continuous copper layer, as can be seen in Fig.~S1 of the Supplementary Material.

\begin{figure}
\centering
\includegraphics{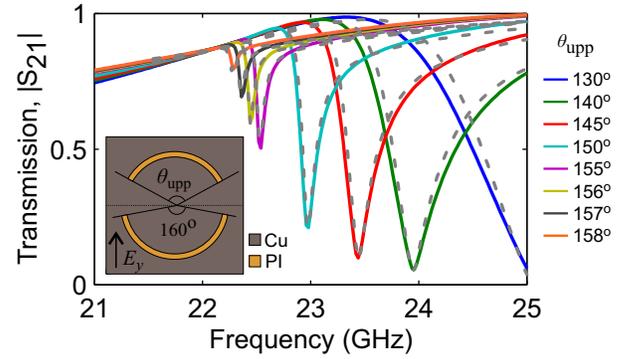}
\caption{Complementary structure: annular slots on an otherwise continuous metallic sheet. Co-polarized transmission coefficient ($S_\mathrm{21}$) under $y$ linear polarization excitation for different $\theta_\mathrm{upp}$ values. The reflection curves of the initial ``metal sectors'' structure under $x$ linear polarization are superimposed after imposing a universal +550~MHz frequency shift. Apart from this $\sim2.5\%$ shift, Babinet's principle is seen to hold well, as expected due to the thin metal and substrate layers.}
\label{fig:Complementary}
\end{figure}

In conclusion, we have theoretically and experimentally studied broken-symmetry metasurfaces in the bottom mm-wave band. The adoption of an ultrathin polymeric substrate allows to minimize resistive losses and obtain high total quality factors, which have been verified experimentally. Both structures (``metal sectors'' and ``annular slots'') can be used to obtain controllably-sharp spectral response, responding to different linear polarizations and offering different characteristics in transmission and reflection.  More generally, a generic platform for low-cost, large-scale engineering of metasurfaces with minimal substrate-induced effects has been identified that can prove useful in broad range of phenomena and applications.

\section*{Supplementary Material}
See supplementary material for additional photos of samples, extended Fano formula for extracting the quality factor from transmission spectra, mode profiles for (i) bright mode of “metal sectors” metasurface and (ii) quasi-dark mode of “annular slots” metasurface, and experimental verification for “annular slots” metasurface.

\begin{acknowledgments}
Support by the Hellenic Foundation for Research and Innovation (H.F.R.I.) under the ``2nd Call for H.F.R.I. Research Projects to support Post-doctoral Researchers'' (Project Number: 916, PHOTOSURF). The authors acknowledge the European Union COST action CA18223 ``Future communications with higher-symmetric engineered artificial materials'' and are grateful for the support received from the CNR Short Term Mobility program 2021.
\end{acknowledgments}


\section*{Data Availability Statement}
The data that support the findings of this study are available from the corresponding author upon reasonable request.



\end{document}